\documentclass[twocolumn,prb,showpacs]{revtex4}
\usepackage{graphics}

\newcommand{\xy}{\textsl{XY} }
\newcommand{\unitx}{\hat{\mathbf{x}} }
\newcommand{\unity}{\hat{\mathbf{y}} }

\begin{document}

\title{Direct Evidence of the Discontinuous Character of the Kosterlitz-Thouless Jump}
\author {Petter \surname{Minnhagen}}
\affiliation {NORDITA, Blegdamsvej 17, DK 2100, Copenhagen,
Denmark}
\author {Beom Jun \surname{Kim}}
\affiliation{Dept.of Molecular Science and Technology, Ajou
University, Suwon 442-749, Korea}

\begin{abstract}
It is numerically shown that the discontinuous character of the helicity
modulus of the two-dimensional \xy model at the Kosterlitz-Thouless (KT) transition
can be directly related to a higher order derivative
of the free energy without presuming any {\it a priori} knowledge
of the nature of the transition. It is also suggested
that this higher order derivative is of intrinsic interest in that
it gives an additional characteristics of the KT transition which
might be associated with a universal number akin
to the universal value of the helicity modulus at the critical
temperature.
\end{abstract}


\maketitle
\section{Introduction} \label{sec:intro}
The Kosterlitz-Thouless(KT) transition has attracted a steady
interest since its discovery.~\cite{ber,kt} This is both because
its unusual characteristic properties and its applicability to
many systems with two-dimensional (2D) character, like,  e.g.,
superfluid/superconducting films.~\cite{minnhagen_rev} The
critical properties are given by Kosterlitz's renormalization
equations (KRE).\cite{k} A key characteristic feature is the
discontinuous universal superfluid jump to zero at the
transition.~\cite{kn} The generic model for the KT transition is
the two-dimensional \xy model and in this case the discontinuous
jump is associated with the helicity modulus.~\cite{kt,ber} The 2D
\xy model has been the subject of very many computer simulations
directed at verifying or disproving the various characteristics of
the KT transition as given by KRE.~\cite{ki,jkk} Although the
general consensus is that the 2D \xy model does indeed undergo a
KT transition, there have also been claims from simulations that
the predictions from KRE may not be entirely
correct.~\cite{ki,jkk} In particular, these earlier works have
been focused on the question whether or not the divergence of the
correlation length (when the transition is approached from above),
is governed by an essential singularity, as predicted by KRE, or
by a conventional power-law singularity. Some evidence for the
latter possibility was found in Refs.~\onlinecite{jkk,sspl}.
Alternatively, in Ref.~\onlinecite{ki} it was argued that the
divergence is given by an essential singularity which is not
entirely consistent with KRE. All this reflects the difficulty of
verifying the precise nature of the transition through computer
simulations.~\cite{egs}

\begin{figure}
\centering{\resizebox*{!}{6cm}{\includegraphics{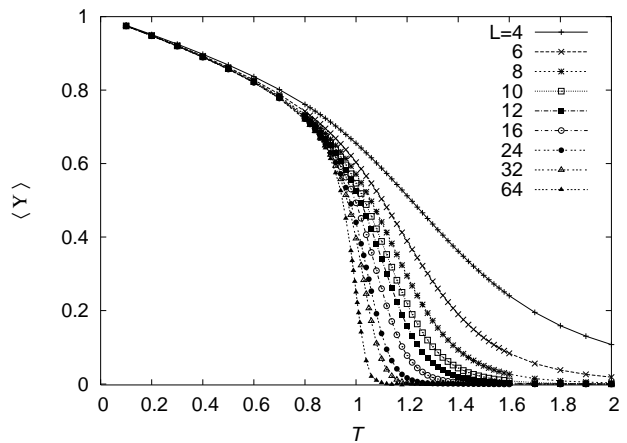}}}
\caption{Helicity modulus
$\langle \Upsilon\rangle $ for the 2D \xy model  for
lattice sizes $L=4$ to 64.
The transition is signaled by a rapid decrease of the helicity
modulus in the vicinity of $T=1$ as temperature is increased. This
decrease becomes sharper with increasing $L$. The question is
whether or not this rapid decrease develops into a discontinuous
jump from a finite positive value to zero as predicted by the
Kosterlitz-Thouless scenario. The data are consistent with such a
scenario, but on the basis of this type of data alone one cannot
rule out that the helicity modulus instead goes continuously to
zero.}
\label{fig:hel}
\end{figure}

Similarly, it is difficult to directly verify from simulations
that there is a  discontinuous jump in the helicity modulus as
predicted by KRE. The reason for this is illustrated in
Fig.~\ref{fig:hel}: Simulations can only be performed on a finite
systems, resulting in the helicity modulus as a continuous
function without any singularity. As seen in Fig.~\ref{fig:hel}
the drop at the transition gets steeper and steeper as the system
size is increased. However, the size dependence is shown to be
rather weak. Consequently, although numerical results like those
presented in Fig.~\ref{fig:hel} for the helicity modulus are
consistent with a discontinuous jump for an infinite system size,
one cannot on this evidence alone, exclude the possibility that
the helicity modulus remains continuous in the thermodynamic
limit. The correctness of the discontinuous character of the jump
has in practice only been verified from simulations in more
indirect ways which make use of additional KRE-predictions for the
KT transition like e.g. that the leading size dependence of the
helicity modulus at the critical temperature is
logarithmic.~\cite{wm}

Thus one may ask if there exists a more direct way of inferring
from simulations that there is a discontinuous jump, without
resorting to more indirect methods which use additional KRE predictions.
In this paper we show that such a more direct way does indeed exist.

Our approach is based on the calculation of a higher order
correlation function. This correlation function appears to give an
additional and somewhat unexpected characteristic feature of the
KT transition. We speculate that this feature might be associated
with a universal number akin to the universal value of the
helicity modulus at the critical temperature.

\section{Model}
The 2D \xy  model on a square lattice of the size $L \times L$
is defined by the Hamiltonian
\begin{equation} \label{eq_H}
 H=- J \sum_{\langle ij\rangle}
\cos\left(\phi_{ij}\equiv\theta_i-\theta_j-
\frac{1}{L}\mathbf{r}_{ij}\cdot \mathbf{\Delta}\right),
\end{equation}
where $J$ is the coupling strength (set to unity from now on),
the sum is over nearest neighbor pairs separated by the
displacement $\mathbf{r}_{ij} = \unitx$ or $\unity$ (we set
the lattice spacing $a \equiv 1$), and
the phase angle $\theta_i$ ($0\leq\theta_i\leq 2\pi$) at the lattice point $i$
satisfies the periodicity $\theta_{i+L\unitx} = \theta_{i+L\unity}
= \theta_i$. For generality, we have also included an externally
imposed global twist across the sample, $\mathbf{ \Delta} = (\Delta_x,
\Delta_y)$, defined by that the summation of the phase difference
$\phi_{ij}$ along the $x$ ($y$) direction equals $\Delta_x$ ($\Delta_y$).
The partition function $Z$ is given by
\begin{equation}
Z=\prod_i\int \frac{d\theta_i}{2\pi}{\rm e}^{-H/T}
\end{equation}
and the free energy by $F({\bf \Delta})=-T\ln Z$. The ground state
corresponds to the configuration where all spins point in the same
direction (i.e., all $\phi_{ij}=0$) which means that the minimum
of the free energy corresponds to ${\bf \Delta}=0$. From the
symmetry in the Hamiltonian, all odd-order derivatives such as
$\partial F/\partial \Delta|_{\Delta=0}$ and $\partial^3
F/\partial \Delta^3|_{\Delta=0}$ vanish. Accordingly, in the
following we will from standard Monte Carlo simulations obtain the
two first nonvanishing derivatives of the free energy with respect
to ${\bf \Delta}$ at ${\bf \Delta}=0$, i.e., the second order
modulus, usually called the helicity modulus,
$\langle\Upsilon\rangle \equiv
\partial^2 F/\partial \Delta^2|_{\Delta=0}$, and the 4th order modulus
$\langle\Upsilon_4\rangle \equiv \partial^4 F/\partial \Delta^4|_{\Delta=0}$,
where $\langle \cdots \rangle$ denotes the thermal average.
The former (the helicity modulus) can be expressed as
\[\langle\Upsilon\rangle=\langle e\rangle-\frac{L^2}{T}\langle s^2\rangle\]
 where
 \[ e \equiv \frac{1}{L^2}\sum_{\langle ij\rangle_x}\cos(\theta_i-\theta_j),\]
\[ s \equiv \frac{1}{L^2}\sum_{\langle ij\rangle_x}\sin(\theta_i-\theta_j), \]
and the sum is over all links in one direction. This means that
$\langle e\rangle$ is the energy per link and $\langle s\rangle$ is the average current per
link in one direction (here taken to be the $x$ direction). Note
that $\langle s\rangle=0$ by symmetry, in contrast to the current-current
correlation $\langle s^2\rangle $ which measures the current fluctuation. The
fourth-order modulus can be expressed as
\begin{equation}\label{Ups_4}
\langle L^2\Upsilon_4\rangle =-4\langle \Upsilon\rangle  + 3\left[
\langle e\rangle -\frac{L^2}{T}\langle (\Upsilon -\langle \Upsilon\rangle )^2\rangle \right]+
\frac{2L^6}{T^3}\langle s^4\rangle
\end{equation}
In the following, we will measure these two correlation functions,
$\langle \Upsilon \rangle$ and $\langle \Upsilon_4 \rangle$, in
standard MC simulations to demonstrate the existence of a
discontinuous jump  of  the helicity modulus to zero at the phase
transition.

\section{Stability Argument} \label{sec:stability}
Since the global minimum of the free energy $F(\Delta)$
corresponds to zero twist, it follows that $F(0)\leq F(\Delta)$. To lowest orders in the $\Delta$ expansion we have
\[F(\Delta)=\langle \Upsilon\rangle \frac{\Delta^2}{2} +\langle \Upsilon_4\rangle \frac{\Delta^4}{4!}\]
This means that $\langle \Upsilon\rangle \geq 0$ since the lowest order
nonvanishing derivative of the free energy will always dominate
for small enough $\Delta$. However, it also implies that the next
order derivative $\langle \Upsilon_4\rangle $ likewise has to be $\geq 0$ at any
$T$ where $\langle \Upsilon\rangle =0$ in the thermodynamic limit. Our argument
is then the simple observation that $\langle \Upsilon\rangle $ cannot go
continuously to zero at the transition temperature $T_c$ if $\langle \Upsilon_4\rangle $ at the same time
approaches a nonzero negative value at $T_c$. But, since
$\langle \Upsilon\rangle $ is indeed zero in the high-temperature phase, this
means that if $\langle \Upsilon_4\rangle $ approaches a negative value at $T_c$
then the jump has to be discontinuous. The point is, as we will
show below, that the conclusion that $\langle \Upsilon_4\rangle $ is nonzero
and negative at $T_c$ can be convincingly drawn from standard
MC simulations.

\section{Simulation Results}
Our simulation of the helicity modulus for 2D \xy model
is shown in Fig.~\ref{fig:hel} up to sizes $L=64$.
The conclusions which can safely
be drawn by analyzing the finite size dependence of $\langle \Upsilon \rangle$
is that it is finite in the low temperature
phase and is zero in the high temperature phase. Since this is
well established we will not discuss it further here. As mentioned
in Sec.~\ref{sec:intro}, the difficult part from a simulation aspect
is to determine {\em how} the helicity modulus approaches zero at
$T_c$ in the thermodynamic limit.

\begin{figure}
\centering{\resizebox*{!}{6cm}{\includegraphics{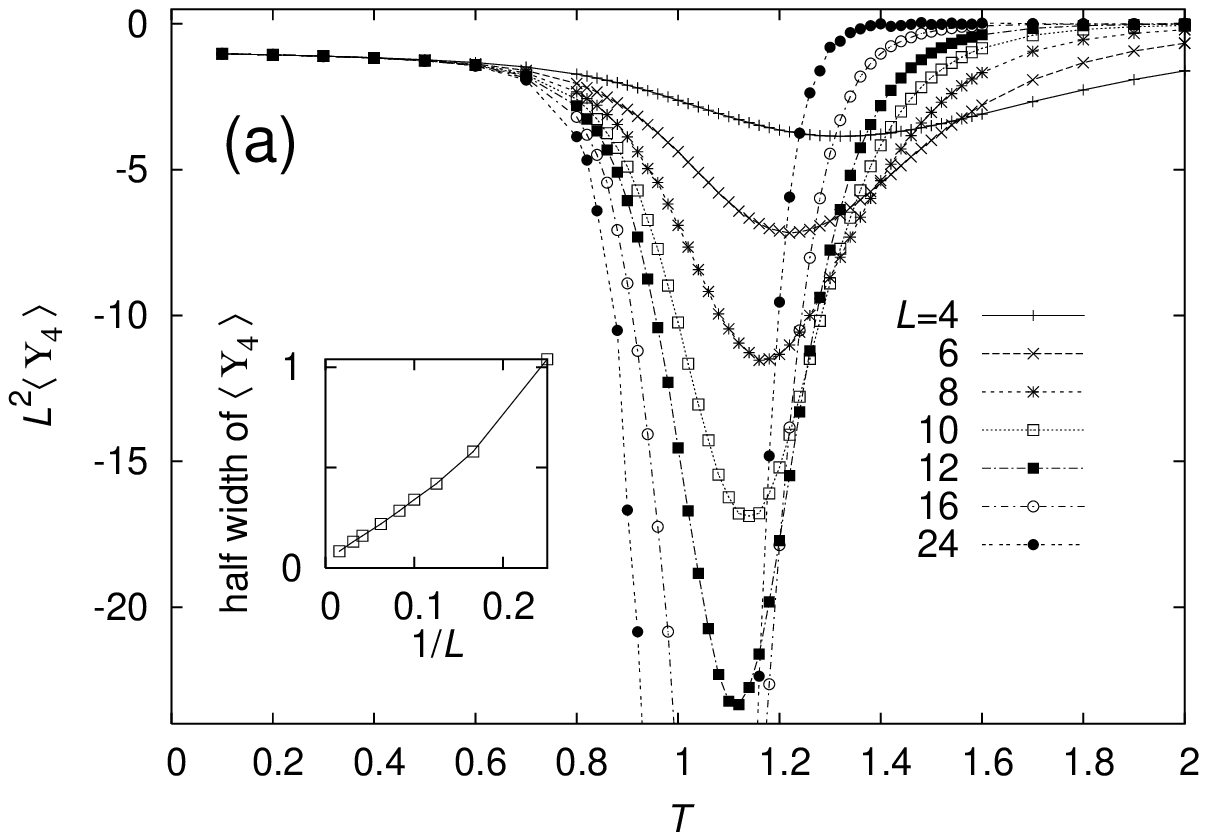}}}

\centering{\resizebox*{!}{6cm}{\includegraphics{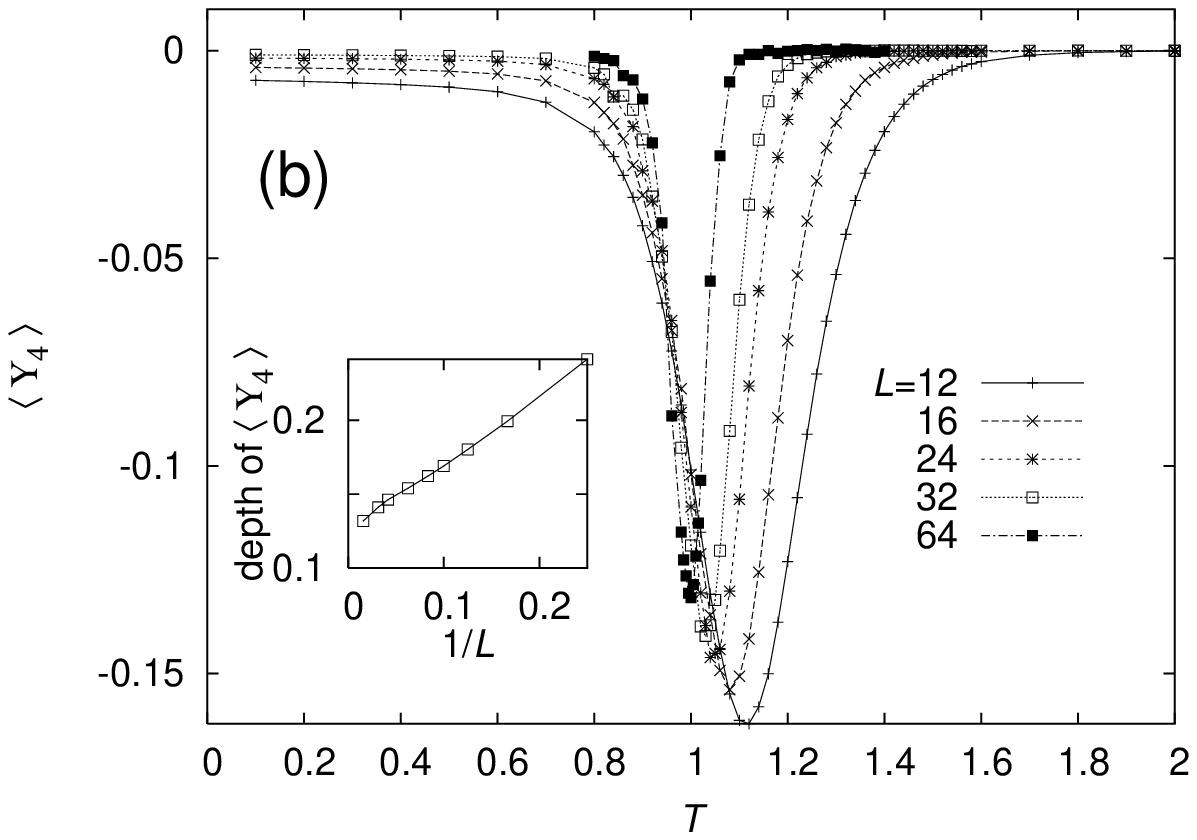}}}
\caption{(a) The correlation function
$\langle L^2\Upsilon_4\rangle $. As seen this correlation function is always
negative or zero. In the high-$T$ phase it approaches zero from
below. In the low-$T$ phase it goes to finite negative values,
 as is apparent from the data points somewhat below $T=1$.
The most interesting feature is the divergent dip in the vicinity
of $T=1$. This divergence at $T_c$ is the manifestation of the
phase transition for this correlation function. The inset
illustrates that the half-width of the dip goes to  zero for large
$L$. (b) The 4th order modulus $\langle \Upsilon_4\rangle $ for
lattices sizes $L=12$ to 64. As seen this modulus goes to zero
both below and above $T_c$. However, the crucial point is that the
depth of the dip remains finite, as is shown in the inset where
the data are plotted against $1/L$. The linear extrapolation to
$L=\infty$ gives $\langle \Upsilon_4\rangle \approx -0.130\pm
0.005$ at $T_c$}. \label{fig:hel4}
\end{figure}

To this end we will use the stability argument given above and
instead focus on the size dependence of the correlation function
$\langle L^2\Upsilon_4\rangle $ (see Fig.~\ref{fig:hel4}).
Figure~\ref{fig:hel4}(a) shows that $\langle L^2\Upsilon_4\rangle
$ vanishes in the high-temperature phase and goes to a nonzero
negative value in the low temperature phase as the system size
becomes larger.  The interesting features is the divergent dip.
This occurs in the region where the helicity modulus for finite
$L$ goes rapidly towards zero. Thus this singularity in $\langle
L^2\Upsilon_4\rangle $ can safely be associated with $T_c$. The
conclusion from Fig.~\ref{fig:hel4}(a) is that $\langle
L^2\Upsilon_4\rangle $ diverges at $T_c$ for $L\rightarrow
\infty$, goes to zero above, and goes to a negative nonzero value
below. This conclusion is further supported by the fact that the
half-width of the divergent dip decreases towards zero for
$L\rightarrow \infty$ as illustrated  in the inset of
Fig.~\ref{fig:hel4}(a). The crucial point is now what this
divergence in the correlation function $\langle
L^2\Upsilon_4\rangle $ implies for $\langle \Upsilon_4\rangle $.
This is shown in Fig.~\ref{fig:hel4}(b), which gives the result
for $L\geq 12$. The inset in Fig.~\ref{fig:hel4}(b) gives the depth
of the dip as a function of size. As is apparent from the inset,
the depth goes to a finite value in the thermodynamic limit.
Linear extrapolation in $1/L$ to $L=\infty$ gives the value
$0.130 \pm 0.005$. Thus we conclude that $\langle \Upsilon_4\rangle
$ is indeed negative and nonzero precisely at $T_c$. Using the
stability argument in Sec.~\ref{sec:stability} this means that the
helicity modulus has to be positive and nonzero precisely at
$T_c$. However, above $T_c$ the helicity modulus is zero from
which follows that the helicity modulus has to be discontinuous at
$T_c$.

One may also note from Fig.~\ref{fig:hel4}(b) that the position of
the minimum of the dip decreases towards lower $T$ with increasing
$L$. However, the approach towards the known value of $T_c$
($\approx 0.89$) is rather slow making a precise determination of
$T_c$ based on $\langle \Upsilon_4\rangle $-data less
advantageous.

\section{Final Remarks}
We have shown that the discontinuous character of the jump of the
helicity modulus at the transition for the 2D \xy model can be
established from MC simulations of the 4th order modulus given by
Eq.(\ref{Ups_4}). It seems very likely that the singularity found
here for the 4th order modulus in case of the 2D \xy model is part
of the general KT transition scenario. Thus we suggest that it
sometimes may be advantageous to study the singularity of this 4th
order modulus in simulations, when trying to determine if a
transition for a particular model is of KT type. It is also
tempting to speculate that the finite value of $\langle
\Upsilon_4\rangle\approx -0.130 $ at $T_c$ is associated with a
universal number, akin to the universal value of of the helicity
modulus $\langle \Upsilon\rangle =2T_c/\pi$.

\section*{Acknowledgements}
Support from the Swedish Natural Research Council through Contract
No.\ F 5102-659/2001 is gratefully acknowledged. B.J.K. was
supported by the Korea Science and Engineering Foundation through Grant
No. R14-2002-062-01000-0.

\end{document}